\documentstyle[aps,prabib,preprint]{revtex}

\newcommand  {\lto }    {\,\vcenter{\hbox{$\buildrel\textstyle<\over\sim$}}\,}

\begin{document}
\title{Self-Pulsating Semiconductor Lasers: 
       Theory and Experiment}

\author{
        C. R.~Mirasso$^1$, 
        G.H.M. van~Tartwijk$^{2,*}$, 
        E.~Hern\'{a}ndez-Garc\'{\i}a$^{3,1}$,
        D.~Lenstra$^{2}$, 
        S.~Lynch$^{4}$, 
        P.~Landais$^{4}$, 
        P.~Phelan $^{5}$, 
        J.~O'Gorman$^{4}$, 
        M.~San Miguel$^{3,1}$, 
        and W.~Els{\"a}{\ss}er$^{6}$
       }
\address{$^{1}$ Departament de F\'{\i}sica, Universitat de les Illes   
                Balears, E-07071 Palma de Mallorca, Spain.\\         
         $^{2}$ Department of Physics and Astronomy, Vrije  
                Universiteit, De Boelelaan 1081, \\ 1081 HV Amsterdam, 
                The   Netherlands.\\         
         $^{3}$ Instituto Mediterraneo de Estudios Avanzados 
                IMEDEA, CSIC-UIB, \\ E-07071  Palma de Mallorca, Spain. \\ 
         $^{4}$ Optronics Ireland, Physics Department, Trinity 
                College, Dublin 2, Republic of Ireland.\\         
         $^{5}$ Physics Department, Trinity 
                College, Dublin 2, Republic of Ireland.\\         
         $^{6}$ Institut f{\"u}r Angewandte Physik,  Technische 
                Universit{\"a}t Darmstadt, Schlo{\ss}gartenstra{\ss}e 
                7, 62289 Darmstadt, Germany.\\   
    $^{*}$   Present address: Philips Optoelectronics B.V,  
    Prof. Holstlaan 4,    5656 AA Eindhoven, \\ The Netherlands
        }
\date{Published in IEEE J. Quantum Electron. 35, 764-770 (1999).}
\maketitle

\begin{abstract}
We report detailed measurements of the pump--current dependency of 
the self-pulsating frequency of semiconductor CD lasers. A distinct 
kink in this dependence is found and explained using rate--equation 
model. The kink denotes a transition 
between a region where the self--pulsations are weakly sustained 
relaxation oscillations and a region where Q--switching takes
place. Simulations show that spontaneous emission noise 
plays a crucial role for the cross--over.
\end{abstract}
\pagebreak

\section{Introduction}
Self-pulsating semiconductor lasers (SPSL's) are of great interest 
owing to their potential application in telecommunication systems as 
well as in optical data storage applications. In particular, in the latter case 
they are realized as so-called narrow--stripe geometry CD lasers 
where the self-pulsation is achieved via saturable absorption in the 
transverse dimension limiting the active 
region. A profound knowledge and understanding of their operation 
dynamics is therefore desired. 

SPSL's have been studied since the first diode lasers became 
available in the late 1960s \cite{Basov68}. These first semiconductor 
lasers, although designed to operate in continuous wave (CW) mode, 
showed self-induced 
pulsations of the light intensity due to a combination of two 
reasons: (i) the laser resonance is internally excited through the 
nonlinear interaction of various longitudinal laser modes, thus 
causing mode beating at very high frequency; 
(ii) defects in the active material act as 
saturable absorbing areas, thus causing absorptive Q-switching 
processes. 

In the case of self-pulsations caused by saturable-absorbing effects, 
the self-pulsation frequency (SPF) dependence on the pump current was
investigated in \cite{dixon}. In later works  
the self pulsations were attributed to undamped relaxation oscillations (RO) 
\cite{Yamada93,Tanaka98}. The precise values of the ROF, as 
calculated from a small-signal analysis, and the actual SPF, highly 
nonlinear, are however different, the SPF being always smaller than the ROF 
\cite{GuidoMaxi}. 

Saturable absorption effects, causing 
self-pulsations in stripe-geometry lasers have been investigated 
since the early 1980s \cite{Paoli79,Ueno85,Mike}. Saturable 
absorption is also responsible for self-pulsations in double-section 
laser diodes \cite{Avrutin93}. A similar mechanism of dispersive 
Q-switching has been invoked to describe self-pulsations in 
multisection Distributed Feedback Lasers 
\cite{Bandelow93,BandelowPREP,Bandelow3}.

In this paper we study both experimentally and theoretically the 
dependence of the self-pulsation frequency (SPF) of narrow--stripe 
geometry self-pulsating semiconductor lasers, also known as CD--lasers,
on the bias pump current.  
In these lasers, self--pulsation is induced via saturable 
absorption in the transverse dimension of the active region. 
The rate--equation model of Ref.\   \cite{Yamada93} has been proven to 
be quite successful in describing the mechanism of self-pulsation and 
has already been used with success in analyzing such lasers subject to 
weak optical feedback \cite{GuidoMaxi}. There it was found that, 
with and without feedback,  
there are two distinct regions in the SPF vs.\  pump--current curve, 
one where spontaneous emission dominates the laser dynamics between 
pulses and one where spontaneous emission always plays a minor role.

In section II we present detailed measurements of the SPF vs
pump--current curve. This curve confirms most of the findings of
\cite{GuidoMaxi}, and also shows a distinct cross-over
point distinguishing between linear and square--root--like behavior.
In section III we confront the experimental results with a 
theoretical model, inspired by Ref.\  \cite{Yamada93}.  
Its results agree qualitatively well with the 
experimental results, showing a distinct cross-over region. 
The location of the cross-over region is shown to be determined
by the spontaneous emission rate. 
In Section IV we discuss the relationship between the SPF and the ROF
using a small signal analysis. We discuss the various bifurcations
that are predicted by our model, and compare it with the   
model of Ref.\  \cite{GuidoMaxi}.

\section{Experiment}

We use a SHARP CD semiconductor laser diode, model LTO22MD. The laser 
emits a continuous train of regular pulses with a frequency that 
depends on the bias pump 
current. A bulk layer of AlGaAs constitutes the active layer 
of this Fabry-Perot cavity that emits at $800$--nm wavelength. 
The gain section is defined by the 
p-electrical contact and has the following approximate dimensions: 
$250$--$\mu$m long, $2$--$\mu$m wide, and $0.2$--$\mu$m thick. A very 
narrow contact of $\sim 2 \mu$m allows for current injection. Since 
the region capable of stimulated emission extends to both sides 
beyond the narrow stripe of the current contact, the wings of the 
optical field distribution will interact with these unpumped, and 
therefore absorbing, regions. In fact, these regions are saturably 
absorbing; when the optical intensity in the wings of the mode is 
large enough, the electron--hole pair population in the unpumped 
region reaches transparency, thus allowing a ``self--Q--switched'' 
pulse.  There is no sharp boundary between the pumped and unpumped 
regions, making carrier diffusion an important effect. 
Indeed, in the model of Ref.\  \cite{Yamada93} carrier diffusion 
between the pumped and unpumped regions is crucial for the appearance 
of self--pulsation. 

The experimental set-up is illustrated in Fig.\  1. The laser is 
temperature controlled by a Peltier cooler at 20 C. It is DC biased 
by a low noise current supply. Laser emission is collected by an 
anti-reflection coated $0.65$ N.A.\  laser diode lens. The resulting 
parallel beam is passed through  a $30$--dB isolator to avoid spurious 
effects caused by optical feedback and is launched into a $60$--GHz 
photodiode (New Focus Model 1006). The converted electrical signal is 
observed with a 
$22$--GHz bandwidth spectrum analyzer (HP 8563A). The typical RF 
spectrum of the SPSL is characterized by a main peak at the SPF, 
followed by overtones. The uncertainty on the self-pulsation 
frequency measurement is mainly due to the measurement of bias 
current, which has an error of less than $0.1$ mA. The resolution of 
the spectrum analyzer is $100$ kHz and the video filter is $30$ kHz.

For values near the threshold current, the low power 
emission makes it difficult to observe the signal. The value of the 
spectral density of the self pulsations is very close to the noise 
level and also the width of the feature in the power spectrum is 
wider than at higher currents. 
To overcome this problem, a small 
current modulation is applied to the device for injection currents below 47 mA
\cite{dublin1},\cite{dublin2}. 
Its power is kept sufficiently low so
that it does not affect the oscillation behaviour of the laser and does not
induce any supplementary oscillation phenomena, e.g. relaxation oscillation
or self-pulsations originating from a cross modulation of the carrier
density. The self-pulsation frequency shows up as an enhancement of the
oscillation of the laser emission if the two frequencies coincide. This
allows an accurate determination of the SPF frequency close to
threshold.

Figure 2 a shows the optical power and SPF as a 
function of the bias current. The L-I curve has been recorded using 
an integrating sphere. It is assumed that all emitted power is 
collected. The laser is characterized by a threshold current of 
$\sim 44$ mA and a slope efficiency of $0.22$ mW/mA. The 
SPF varies from $1$ to $4$ GHz in a bias current range of $46$ to $64$ mA, 
which was the maximum injection current we could reach with these
devices. 
In the region of the lasing threshold the experimental values
present a square-root like behavior dependence reminiscent of 
standard relaxation oscillations as exhibited by a CW-semiconductor laser.
For bias currents above 55 mA this dependence 
was no longer observed and the SP behavior appears to have a more 
linear dependence on the bias current.  

\section{Theory}

In this section we use a simple model to explain the observed 
bias--current dependence of the SPF. The investigated laser has 
a narrow--stripe geometry, which can be modeled in a straightforward 
way using rate-equations \cite{Yamada93} for the optical intensity 
$S$ (suitably normalized to represent the number of photons in the 
cavity), the number of electron--hole pairs $N_1$ in the pumped 
region, and the number of electron--hole pairs $N_2$ in the 
unpumped (absorbing) region:
\begin{mathletters}
\begin{eqnarray}
\frac{dS}{dt}&=&[g_1(N_1-N_{t1})+g_2(N_2-N_{t2})-\kappa] S
    + R_{sp} + F_S(t), \label{2ndSdot}\\ 
\frac{dN_1}{dt} & = & \frac{J}{e}-\frac{N_1}{\tau_s}   
                  -g_1 (N_1-N_{t1})S                     
                  -\frac{N_1-v N_2}{T_{12}} , \label{2ndN1dot} \\
\frac{dN_2}{dt} & = & -\frac{N_2}{\tau_s}
                  -g_2 (N_2-N_{t2})S                      
                  +\frac{N_1/v- N_2}{T_{21}}. \label{2ndN2dot} 
\end{eqnarray}
\end{mathletters} 
where $g_1$ ($g_2$) is the gain coefficient at the transparency number 
$N_{t1}$ ($N_{t2}$) in the pumped (unpumped) region, $\kappa$ is the 
total loss rate. $R_{sp}=\beta_{sp} \eta_{sp} N_1 / \tau_s$ is the spontaneous 
emission rate, $\eta_{sp}$ is the spontaneous quantum efficiency, $\beta_{sp}$ is
the spontaneous emission factor and $\tau_s$ is the carrier lifetime. 
$F_S(t)$ is a delta-correlated Langevin noise source \cite{Henry82} with  
correlation  
$<F_S(t_1)F_S(t_2)> = 2 R_{sp} S \delta (t_1-t_2)/\tau_s$, 
$J$ is the bias pump-current, $e$ is the 
elementary charge,  
$v=V_1/V_2$ is the volume ratio of pumped and unpumped region, 
$T_{12}$ is the diffusion time from the pumped region to the 
unpumped region, and $T_{21}$ is the diffusion time from unpumped to 
pumped region. These two diffusion times are interrelated through the 
volume ratio $v$ \cite{Yamada93,GuidoMaxi}:
\begin{equation}v = \frac{V_1}{V_2} = \frac{T_{12}}{T_{21}} .
\label{T12def}
\end{equation}
Our model (\ref{2ndSdot}-\ref{2ndN2dot}) is a simplification of
the model used in Ref.\  \cite{GuidoMaxi}, where the carrier
dependence of the carrier lifetime $\tau_s$ is taken into account 
using the well-known second order expression for $\tau_{s}^{-1}$ in 
the carrier number $N_j$. Here, we neglect this dependence for the 
moment, as it simplifies the analytical work and qualitatively gives
similar results.

Using the parameter values listed in Table 1, Eqs.\ 
(\ref{2ndSdot}-\ref{2ndN2dot}) are numerically solved with a standard 
algorithm \cite{GuidoMaxi}. In Figure 
3 we show the resulting SPF-J curves, with and without
spontaneous emission noise. 
Each value of the curves is calculated from an average over $10^3$ pulses. 
It is seen that the observed kink in the SPF-J curve
is the result of spontaneous emission noise. There is a shift of the kink 
towards larger currents upon increasing the spontaneous emission level. For the
values of table 1 and $\beta_{sp}=1.3 \times 10^{-6}$, $J_{xover} \approx 82$ 
mA. It should be noticed that we do not expect a quantitave agreement between
experimental and numerical results, since the model neglects important 
effects, such as gain saturation. Nevertheless, the qualitative trends are
well reproduced allowing us to physically understand the origin of the 
experimental features.

Figure 4 shows time traces of the intensity for different bias currents.
Clearly, the interpulse 
intensity drastically increases with current in the vicinity of the 
kink. For currents $J<< J_{xover}$ the interpulse intensity is well 
dominated by the spontaneous emission (panel a)), while for currents 
$J>J_{xover}$ spontaneous emission does not affect the intensity 
significantly. 
The kink-current $J_{xover}$ can be defined as the highest 
current at which the interpulse intensity is dominated by spontaneous
emission noise.  As can be seen in Eq.\  (\ref{2ndSdot})
spontaneous emission increases the intensity generation rate 
with an amount $R_{sp}$. The effect of this 
on the self--pulsation process depends on the generation rate
through stimulated emission $R_{stim} = 
[g_1(N_1-N_{t1}) + g_2 (N_2-N_{t2})]S$. For currents 
$J<J_{xover}$, $R_{sp} > R_{stim}$ in the interpulse region 
while for $J>J_{xover}$, the contrary happens. 
Therefore, the kink pump current $J_{xover}$ could be mathematically
identified through
\begin{equation}
R_{sp} \equiv R_{stim}(J_{xover}), \label{Jxoverdef}
\end{equation}
where the current dependence of $R_{stim}$ reflects the 
need to solve Eq.\  (\ref{Jxoverdef})  
implicitly using all three equations (\ref{2ndSdot}-\ref{2ndN2dot}) at the time
at which the intensity reaches the minimum.

The long--dashed curve in Fig.\  3 is obtained by putting $R_{sp}=0$
in Eq.\ (\ref{2ndSdot}). In that situation, the interpulse intensity 
becomes extremely small upon decreasing the pump current. The
smaller the interpulse intensity becomes, the longer it takes for
the absorber to reach transparency. When including noise ($R_{sp}\neq
0$), the interpulse intensity remains at a much higher level in the 
same pump current interval because of the spontaneous
emission rate $R_{sp}$. This will significantly increase the 
speed with which a new pulse is generated after the previous 
one has depleted the absorber. We note that the Langevin 
noise source $F_S(t)$ in Eq.\  (\ref{2ndSdot}) is responsible for
the timing jitter of the pulses. In the region $J<J_{xover}$ a single 
noise event in between pulses may significantly delay or advance the
birth of the next pulse, causing substantial jitter. For pump 
currents above the cross--over, the relative effect of the noisy 
events, and hence the jitter, is much smaller. The existence of
two pump current regions with very different jitter characteristics
was also found in Ref.\  \cite{GuidoMaxi}.

In figure 5 the maximum pulse intensity ($S_{max}$) and the minimum interpulse 
intensity ($S_{min}$) vs. the bias current are shown. An abrupt change (note
that the scale in panel c) is logarithmic) of $S_{min}$ can be seen 
at $J_{xover}$ (while $S_{max}$ takes it maximum value).
The kink-current $J_{xover}$ is therefore identified as the highest 
current at which the interpulse intensity is dominated by spontaneous
emission noise. The kink also denotes the boundary between two regimes 
that can be 
described as follows: For currents larger than $J_{xover}$ 
the self-pulsation has the character of undamped RO, while for 
currents below this value clear self-Q-switching takes place. 
Obviously, for currents $J>J_{xover}$ the absorber is not depleted 
deeply enough to cause a Q-switch: as soon as transparency is 
reached, the absorber is bleached but the pump is strong enough to 
prevent total bleaching. For currents 
$J<J_{xover}$, the pump is small enough to allow total bleaching of 
the absorbing regions, after which the number of electron--hole pairs 
in the absorbing region has to start all over again. No bifurcation 
in the usual sense can, however, be attributed to this critical 
current. 

In the next section, we will look at the relationship between 
ROF and SPF in more detail. 

\section{Relaxation Oscillations and Self-Pulsations}

In the previous section we introduced a simple model which provides
an explanation for the peculiar cross-over region in terms of
the average level of spontaneous emission. Here we will put our 
numerical findings in an analytical framework, which leads to 
a clearer picture of the self-pulsation characteristics.

This is achieved by solving for the CW solutions of Eqs.\ 
(\ref{2ndSdot}-\ref{2ndN2dot}) and investigating their stability 
properties. 
First we look for laser threshold, which is defined as the 
circumstance for which the trivial solution ($S=0$) looses stability
in the absence of spontaneous emission.
We therefore put $R_{sp}=0$ in Eq.\  (\ref{2ndSdot}) and obtain:
\begin{eqnarray}
N_{th}&=&\frac{g_1\,N_{t1}~+~g_2\,N_{t2}~+~\kappa}
                 {g_1~+~\frac{g_2\,\tau_s}
                             {T_{12}~+~v\,\tau_s}
                 } \label{Nth} \\
\frac{J_{th}}{e} &=& N_{th}~[\frac{1}{\tau_s}~+~\frac{1}{T_{12}}
        ~-~\frac{v}{T_{12}}~ \frac{\tau_s}{T_{12}~+~v \, \tau_s}] 
       \label{thresh}
\end{eqnarray}
Using the parameters listed in Table 1, we find $J_{th} = 44.53$ mA.

In total, Eqs.\  (\ref{2ndSdot}-\ref{2ndN2dot}) have three possible 
CW solutions. 
Below threshold, only the solution with $S=0$ is physically meaningful
(the other two have negative power). At threshold the solution $S=0$ 
becomes unstable while one of the other two becomes stable with 
positive power. This is found after performing
a standard linear stability analysis, which yields for every CW 
solution a set of (complex) characteristic exponents 
$\lambda = \lambda_r + i\lambda_i$. When any of these exponents has
a positive real part ($\lambda_r>0$), the CW solution is unstable. 
The imaginary part $\lambda_i$ denotes the frequency with which 
perturbations {\em initially} will grow.  Figure 6 shows how the 
real parts of the characteristic exponents of the relevant CW solution
vary with bias current. The CW solution is found to be unstable 
on the interval $44.556 \lto J \lto 92$ mA. For bias currents 
$J > 92$ mA, stable CW emission is found. 
On the other side of the interval, a more complex behavior is 
found.
At $J=44.53$ mA, the CW solution is stable, but looses its stability 
already at $J=44.556$ mA. 
This sequence of bifurcations from the nonlasing ($S=0$) state to 
self pulsation occurs in a very narrow range of currents around 
threshold. Thus the sequence will be experimentally very hard to
resolve due to different noise sources; the laser will seemingly begin 
to oscillate as soon as it crosses threshold. 

Thus, our model (\ref{2ndSdot}-\ref{2ndN2dot}) shows that there exists
a CW solution that looses stability at $J=44.556$ mA and regains 
stability at $J=92$ mA. In  between these values, the CW state is
unstable, as indicated by a complex conjugate pair of characteristic
exponents with positive real parts (Hopf-instability). 
The region of instability 
coincides obviously with the region of self-pulsating behavior, and 
is bounded by two Hopf-bifurcations.
When the laser operates at a bias current $44.556 < J < 92$ mA, 
small perturbations to the CW state in question initially grow as
$\exp{[(\lambda_r + i\lambda_i)t]}$, i.e., with angular 
frequency $\lambda_i$. The linear stability 
analysis does not 
provide any information on how this initial growth will saturate.
Numerical results from Eqs.\  (\ref{2ndSdot}-\ref{2ndN2dot}) 
show that the resulting SPF is always smaller than the ROF
$\lambda_i/2\pi$. This is illustrated in Fig.\  3. Both frequencies
meet at $J=44.556$ mA and at $J=92$ mA, the two Hopf bifurcation
points. In the former case it means that the SPF must increase when 
coming from higher bias currents to reach the RO value. However,
this increase only occurs in a very small range of currents so that 
it would be very hard to observe in the experiment.

It should be noted that a different scenario is found in 
Ref.\  \cite{GuidoMaxi}. There, the carrier lifetime $\tau_s$ 
is considered to be carrier dependent, to account for the 
radiative, non-radiative, and Auger processes \cite{AGDUT}:
\begin{equation}
\tau_{s,j}^{-1} (N_{j}) = A_{nr,j} + B_{j}N_j + C_j N_{j}^{2},
\label{Tjjdef}
\end{equation}
where $j= 1$ denotes the pumped region and $j=2$ denotes the 
unpumped region.  This carrier 
dependence is considered necessary because during the strong 
pulsations, large variations in the carrier numbers $N_j$ may occur 
\cite{AGDUT}. 

It was found in Ref.\  \cite{GuidoMaxi} that the carrier dependence 
of $\tau_{s,j}(N_j)$ plays a significant role around threshold.  This 
is in sharp contrast with the well-known CW edge-emitting lasers where 
$N$ is clamped immediately above threshold.  The kink region, 
lying far above threshold, is not affected significantly by taking 
into account the carrier dependence of $\tau_s$. This illustrates the 
robustness of the cross-over behavior. At the high end of the self-pulsation 
interval, 
also a Hopf bifurcation is found, but the dynamics at the low end differs 
from the one discussed here. First of all, 
there is no window of stability just after threshold. Figure 7 shows 
the location of the various CW solutions as a function of bias pump 
current. The $S=0$ solution (horizontal solid line) is only shown for 
currents where it is stable.  It looses stability at $J=J_{th}$.
Around $J=0.85 J_{th}$ two CW solutions are born out of a bifurcation. 
Both CW solutions are linearly unstable. 
This is in contrast with the model discussed above 
where the upper branch CW solution is stable in a short pump interval 
after its birth. The bifurcation which 
starts the self-pulsation is not a Hopf one but a homoclinic 
bifurcation (collision of a limit cycle and saddle).
Thus, in the model of Ref.\  \cite{GuidoMaxi} self-pulsation occurs in a 
region bounded by a Hopf-bifurcation on the high bias side and a 
homoclinic bifurcation at the low bias side. This type is
not uncommon in (passive Q-switching) self-pulsating lasers with saturable 
absorbers \cite{Lugiato78,Erneux81,Arimondo85,deTomasi89,Hennequin}.

\section{Conclusions}

We have investigated, both experimentally and theoretically, 
the dependence of the self-pulsation frequency of semiconductor 
CD lasers upon changes in the bias current. A distinct kink 
is found in this dependence, which is investigated using 
a rate--equation model. We have identified that
the kink is caused by spontaneous emission, whose average intensity 
sets a lower bound on the emitted laser intensity and thereby on the 
average intensity, which determines the relaxation oscillation 
frequency. 

From our analysis we conclude that below the crossover point, the self-pulsations
behave as passive Q-switching oscillations while above the crossover the behavior
approaches undamped relaxation oscillations.

The relationship between the relaxation oscillation frequency and 
the self-pulsation frequency is investigated by means of a small
signal analysis. We observe that the relaxation oscillation frequency so obtained
is an upper limit for the self-pulsating frequency.
It is also found that self-pulsation occurs in a bias 
current interval bounded by two Hopf bifurcations. A small window
of stable CW emission is found very close to the laser threshold in the absence of
spontaneous emission. 
The model of Ref.\  \cite{GuidoMaxi} does not show such a window 
of stable emission terminated by a Hopf bifurcation, but a 
homoclinic bifurcation is responsible for the onset of the 
self-pulsating behavior. However, for the lasers we used in the experiment, such
differences between the models are irrelevant since they occur in a very small
range of currents too close to threshold to be resolved.  
These results raise an interesting question on the nature of the 
bifurcation at the lower side of the self-pulsation interval. 

{\sl Note added:} A bifurcation analysis of the Yamada model 
neglecting interstripe diffusion has been 
published\cite{krauskopf}. The bifurcation scenario is different 
from the one for our model (\ref{2ndSdot})-(\ref{2ndN2dot}) (and 
closer to the one in \cite{GuidoMaxi}). However, these differences 
in the deterministic behavior have no physical relevance, since 
they appear in parameter domains for which the dynamics is 
dominated by noise.  

\section*{Acknowledgments}
This work was supported by the European Union through the Project 
HCM CHRX-CT94-0594.

\pagebreak

\begin{table}
\caption[]{Meanings and values of the parameters in the rate equations}
\vspace{1 cm}
\begin{tabular} {clcc}
Parameter&Meaning&Value&Units\\
&&&\\
$g_1$&Gain parameter of active region&$4.7 \times 10^{-9}$&ps$^{-1}$\\ 
$g_2$&Gain parameter of absorbing region&
                                       $1.5\times 10^{-8}$&ps$^{-1}$\\ 
$\kappa$&Inverse photon lifetime&$0.4$&ps$^{-1}$\\ 
$\tau_s$&Carrier lifetime         &$1.1$&ns\\
$\beta_{sp}$ &Spontaneous emission coefficient&variable&dimensionless\\ 
$\eta_{sp}$ &Spontaneous quantum efficiency &0.33&dimensionless\\ 
$N_{t1}$&Carrier number at transparency (active region)&$6.\ 
10^7$&dimensionless\\$N_{t2}$&Carrier number at transparency 
(absorbing region)&$1.2\ 10^8$&dimensionless\\
$J_{th}$&Threshold current of the solitary laser&$44.53$&mA\\ 
$\alpha$&Linewidth enhancement factor&5&dimensionless \\ 
$v$&Ratio of the active and absorbing volumes&0.115&dimensionless \\ 
$T_{12}$&Diffusion time &2.1&ns
\end{tabular}
\end{table}

\pagebreak

\begin{figure}
\noindent
\caption[]{Experimental set-up.}
\end{figure}

\begin{figure}
\noindent
\caption[]{Experimentally observed bias current dependence
of the emitted laser power (solid line) and the self-pulsating 
frequency (plus signs).}
\end{figure}

\begin{figure}
\noindent
\caption[]{Self-Pulsating Frequency (SPF)  vs. Bias Current, 
obtained by numerically solving 
Eqs.\  (\ref{2ndSdot}-\ref{2ndN2dot}).
The solid line indicates the value of the relaxation oscillation obtained from
the small signal analysis. 
Long--dashed line: SPF in the absence of noise; 
dash-dotted line: SPF with $\beta_{sp} = 1.3 \times 10^{-4}$;
short-dashed line: SPF with $\beta_{sp}= 1.3 \times 10^{-6}$; 
and dashed-triple-dotted line: $\beta_{sp} = 1.3 \times 10^{-8}$.}
\end{figure}

\begin{figure}
\noindent
\caption[]{Time traces of the intensity as a function of the bias 
current: a) $J=45$ mA, b) $J=80$ mA, c) $J=84$ mA. $\beta_{sp}=1.3
\times 10^{-6}$}
\end{figure}

\begin{figure}
\noindent
\caption[]{Self pulsating frequency (top a)), maximum pulse intensity $S_{max}$
(middle b))and minimum interpulse intensity $S_{min}$ (bottom c))
vs. bias current for $\beta_{sp}=1.3 \times 10^{-6}$.}
\end{figure}

\begin{figure}
\caption[]{The real part $\lambda_r$ of the larger characteristic 
exponent of the CW solution with positive intensity as a function of bias
current. The inset, where the real part of all eigenvalues is included, 
shows the tiny window of stable CW emission just after threshold.}
\end{figure}

\begin{figure}
\noindent
\caption[]{The intensity of the three CW solutions as a function 
of bias current, when the carrier dependence of the carrier
lifetime is included.}
\end{figure}

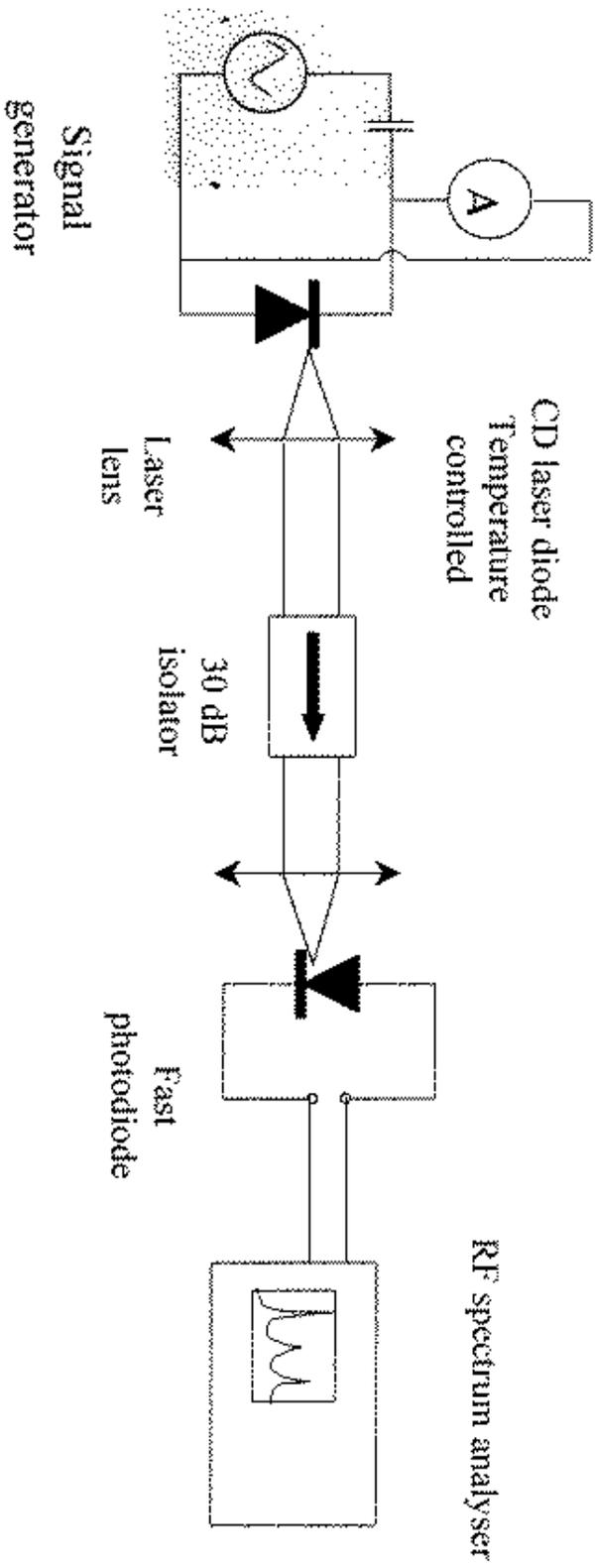
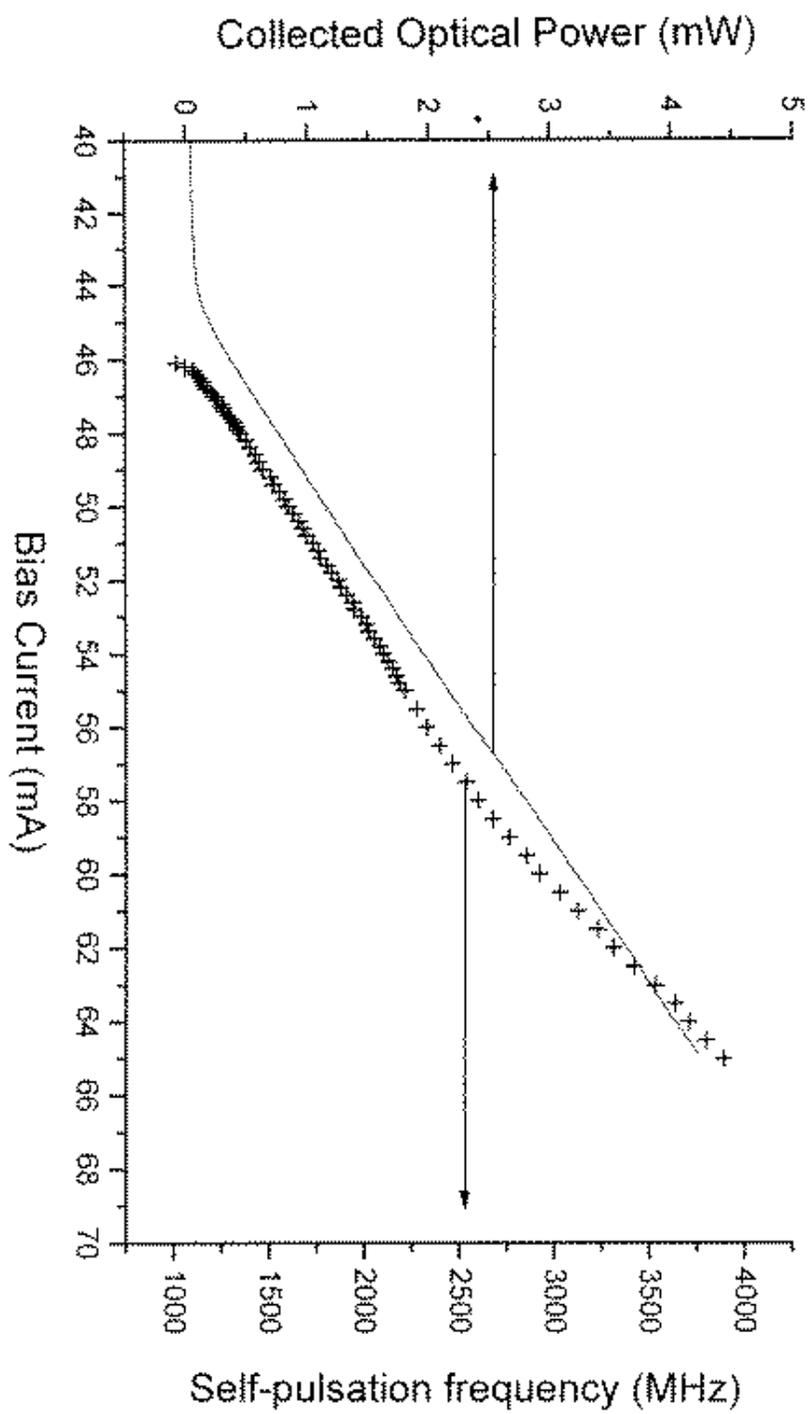
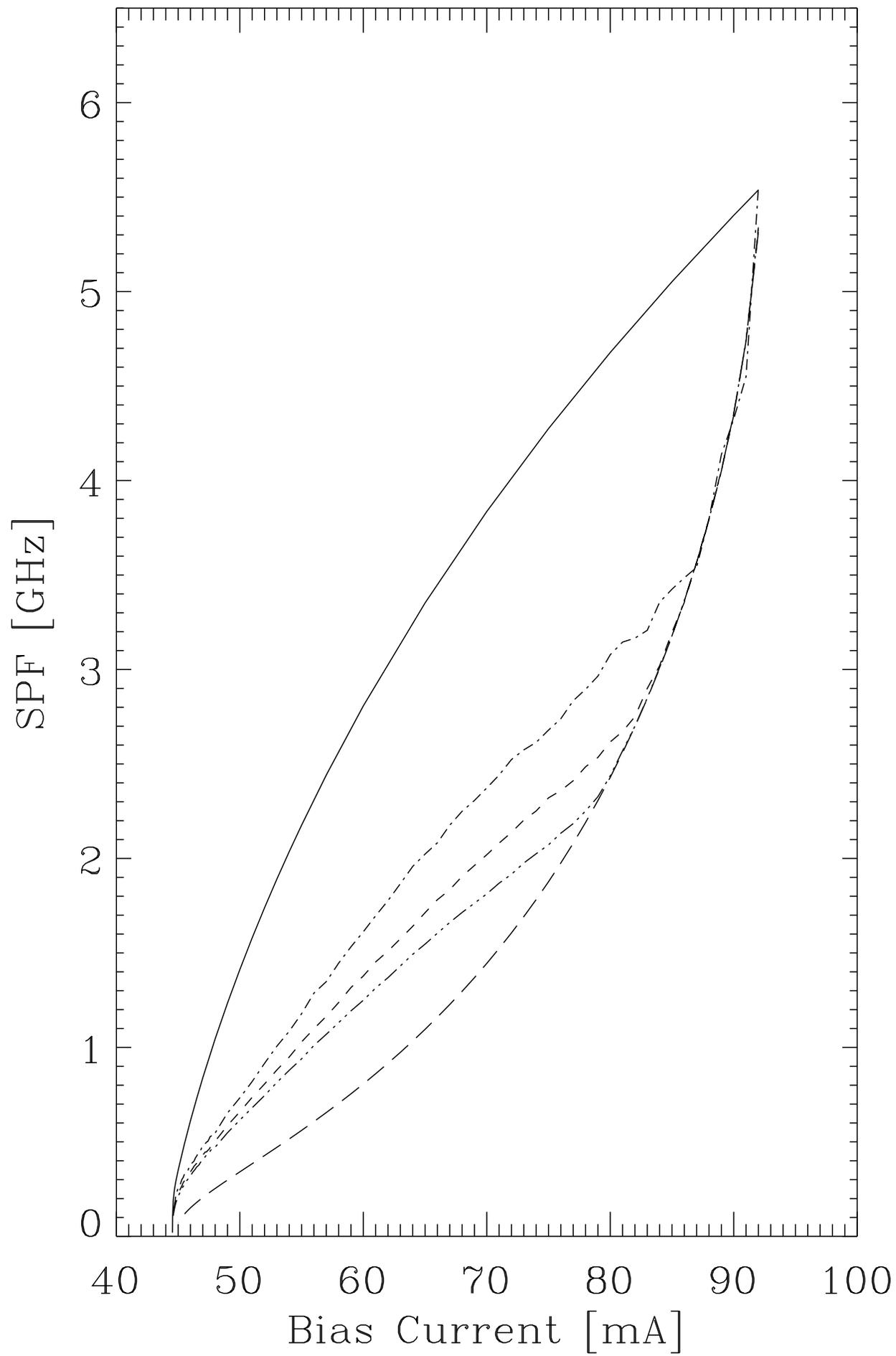
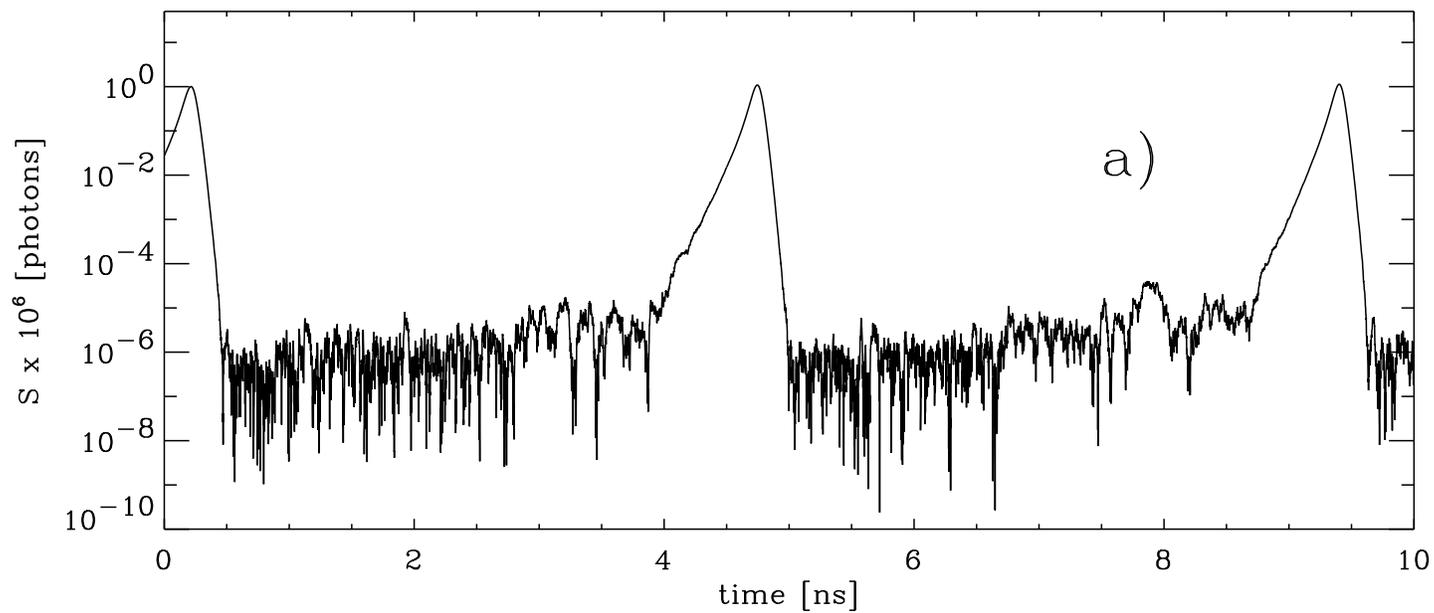
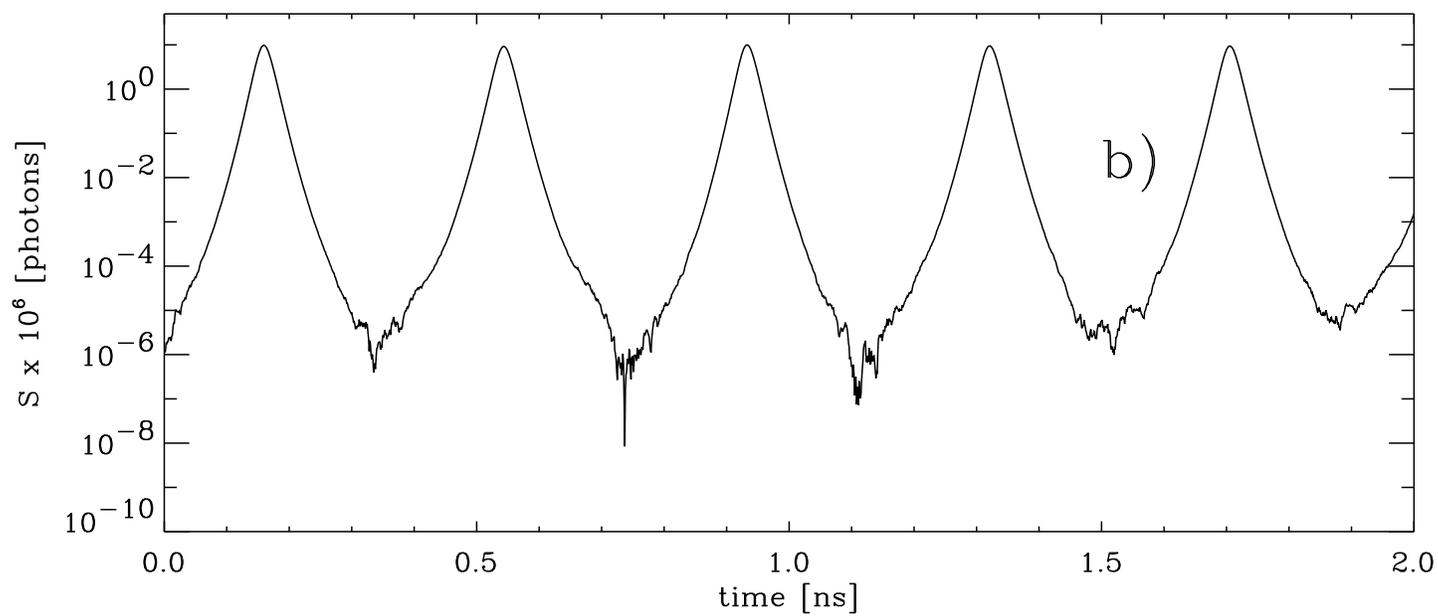
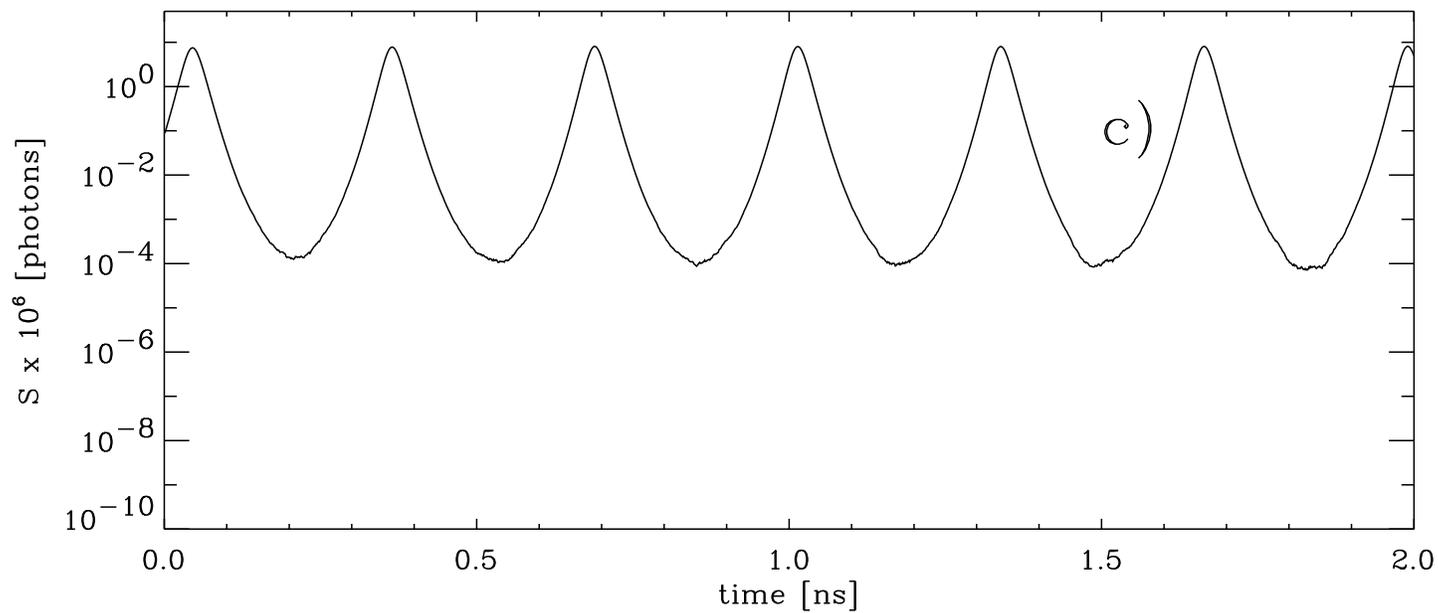
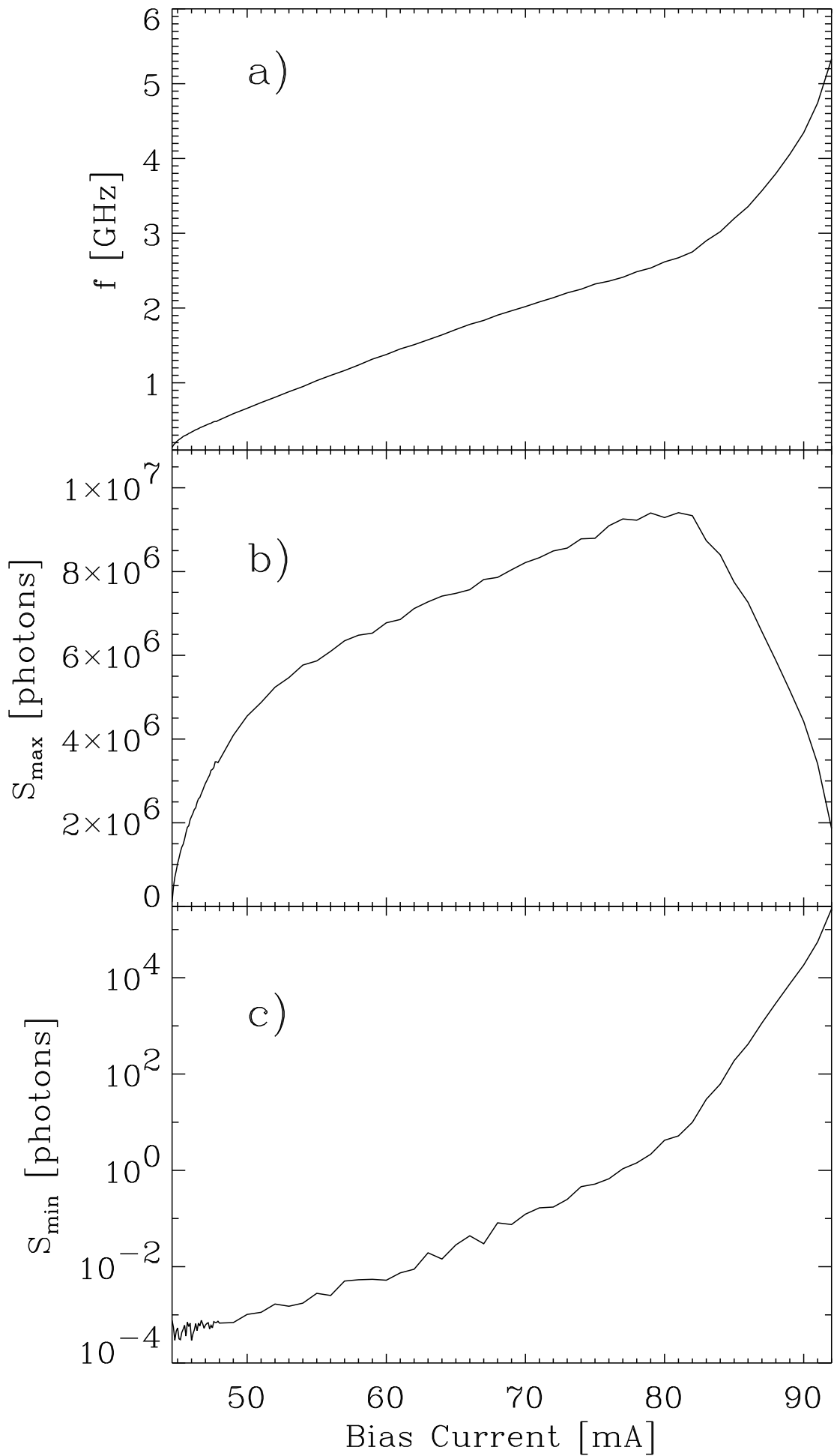
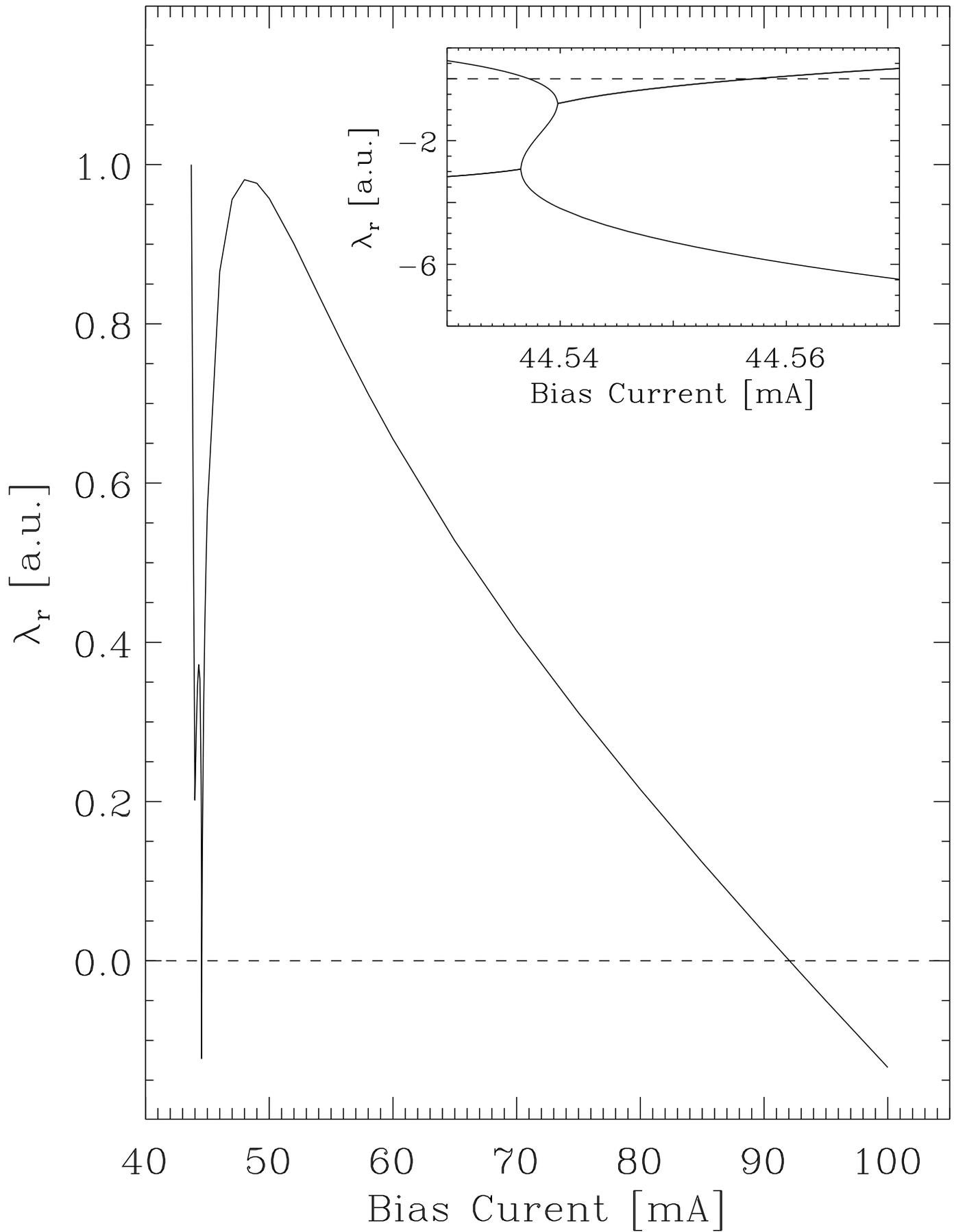
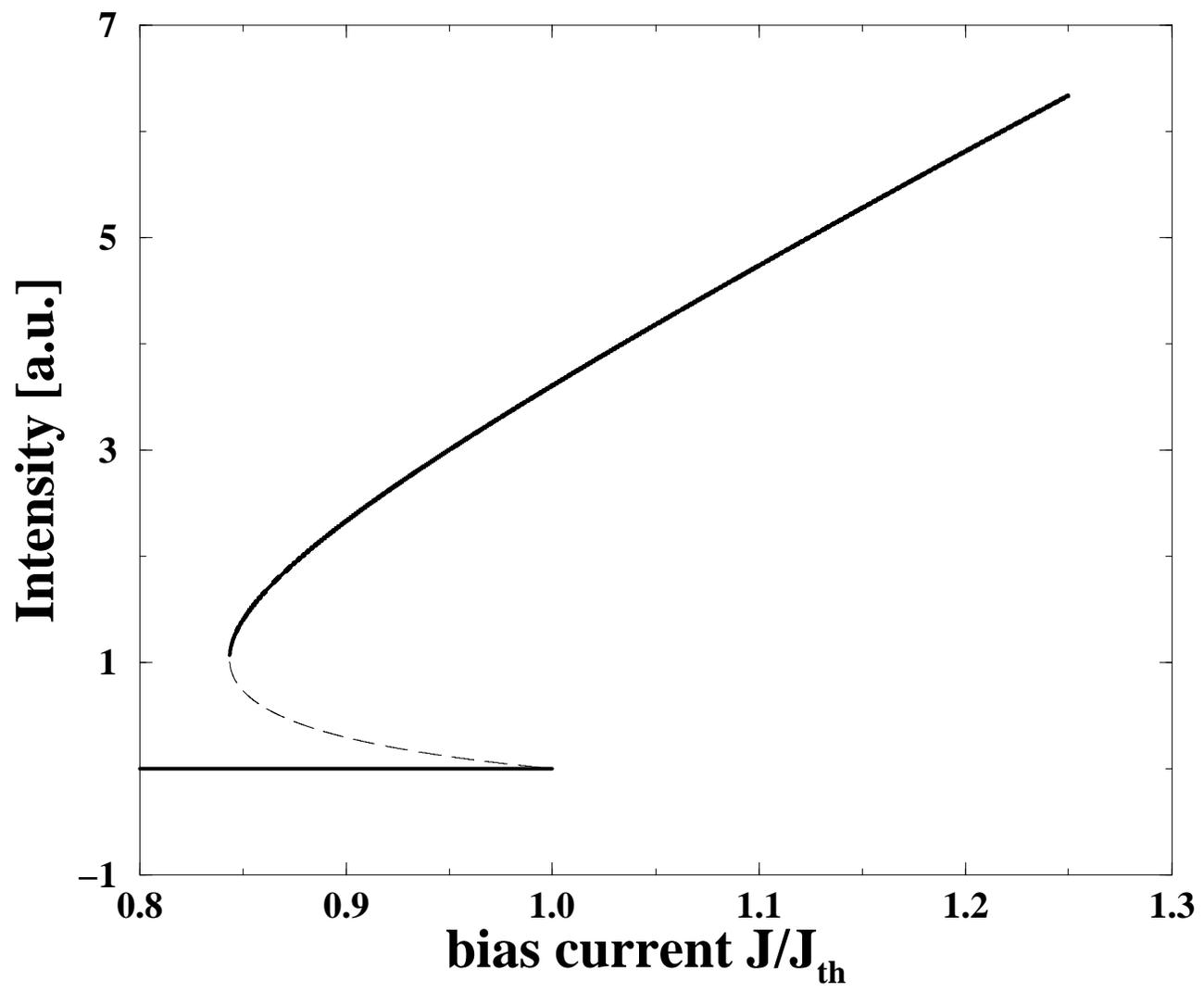
\end{document}